\documentclass[aps,prl,twocolumn,showpacs]{revtex4}

\usepackage{graphicx}
\usepackage{dcolumn}
\usepackage{bm}

\begin{document}

\title{Power-law rheology and mechano-sensing in a cytoskeleton \\
model with forced protein unfolding }


\author{Brenton D. Hoffman}

\author{Gladys Massiera$^\dagger$}

\author{John C. Crocker}

\affiliation{Department of Chemical and Biomolecular Engineering,
University of Pennsylvania, Philadelphia, Pennsylvania 19104, USA}
\date{\today}
\begin{abstract}
We describe a model of cytoskeletal mechanics based on the force-induced
conformational change of protein cross-links in a stressed polymer network.  
Slow deformation of simulated networks containing cross-links that undergo
repeated, serial domain unfolding leads to an unusual state---with many 
cross-links accumulating near the critical force for further unfolding.  Thermal 
activation of these links gives rise to power-law rheology resembling the
previously unexplained mechanical response of living cells.  Moreover, 
we hypothesize that such protein cross-links function as biochemical 
mechano-sensors of cytoskeletal deformation.
\end{abstract}
 \maketitle

The importance of mechanical cues for understanding cell behavior is 
increasingly recognized. Stem cell differentiation \cite{mcbeath}, tissue 
morphogenesis \cite{pascek} as well as cell growth \cite{wang} and 
death \cite{chen} are known to be affected by cell shape or the stiffness 
of the surrounding extra-cellular matrix. The molecular mechanisms by which 
such mechanical cues produce biochemical responses, however, remain 
essentially unknown \cite{janmeyc}. Similarly, the anomalous mechanical
response of cells defies explanation despite the identification of the 
cytoskeleton's major structural constituents and considerable modeling 
effort \cite{gardela,wachsstock,ningwang}.  Rheological measurements on 
living cells yield a frequency dependent shear modulus that scales as a 
weak power-law,  $|G^*(\omega)|\sim\omega^\beta$, with 
$0.1 < \beta < 0.25$, spanning many decades of frequency 
\cite{fabry, feneberg, alcaraz}. Such a mechanical response is rather 
unusual \cite{sollicha}, having only been observed previously in seemingly 
unrelated materials such as foams and pastes.  While networks of purified 
filamentary biopolymers (e.g. F-actin) generally have a frequency-independent 
plateau elasticity \cite{gardelb}, studies of networks containing the 
cross-link proteins filamin \cite{gardel-PNAS,gardelc} 
or $\alpha$-actinin \cite{tseng} have reproduced the weak power-law 
rheology as well as other aspects of the cell response, such as 
stress-induced stiffening.  These results suggest that the power-law rheology
of cells may be due to cross-link conformational change or unbinding. 

In this Letter, we propose a simple cytoskeletal architecture that may 
explain both cell rheology and mechano-sensing, based on the force-induced 
serial unfolding of domains in protein cross-links.  We show, using simulation, 
that shearing a simplified model network leads spontaneously to the 
accumulation of many cross-links at tensions on the cusp of unfolding. 
Thermally activated unfolding of these cross-links readily reproduces 
cells' power-law frequency dependent shear modulus with physically 
plausible model parameters.  Comparable models based on forced 
{\it unbinding} (e.g. of cross-links' actin binding domains from actin) rather
than unfolding do not produce power-law rheology. Moreover, this
unusual, near-critical arrangement of cross-links is suggestive of a 
mechano-sensory function.  We hypothesize that, by modulating the 
binding of signaling species, unfolding cross-link domains function as
the fundamental biochemical transducers of cytoskeletal deformation.

\begin{figure}[t]
\begin{center}
\includegraphics[width=2.75in]{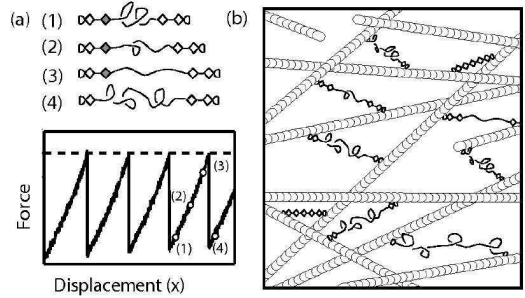} 
\end{center}
\caption{\label{fig.one} Schematic representation of serial unfolding and the
cytoskeleton. (a) As cross-links are extended, domains serially unfold with 
typical force vs. extension curves having abrupt transitions at a critical force, 
$F_c$, corresponding states are labeled. (b) Our model of the cytoskeleton 
consisting of a network of generic semi-flexible polymers and partially 
unfolded, extensible cross-links.}   
\end{figure}

Many cross-linking proteins (e.g. filamin, $\alpha$-actinin, spectrin, 
plakin, etc.)  have a structure consisting of similar, repeated domains 
that can be serially unfolded by applied force \cite{furuike, riefa, riefb},
see Figure \ref{fig.one}(a). The unfolding time is typically described by the 
Bell model: \cite{bell}
\begin{equation} \tau_B(F)=\tau_a
\exp[E_B(1-\frac{F}{F_c})],E_B=\frac{F_c r_o}{k_BT}
 \label{Eq.one}
\end{equation}
where $\tau_a$ is a molecular attempt time, $F_c$ is a critical force, 
$r_o$ is a characteristic bond length-scale, and $k_BT$  is the thermal
energy. The exponential form causes small changes in tension to yield 
rather large changes in unfolding time.  As domains serially unfold, 
the molecule becomes progressively longer and longer.  The entropic 
elasticity of the unfolded protein causes a spring-like response between
unfolding events, producing a `sawtooth' force-extension profile, Figure
\ref{fig.one}a.

We hypothesize that some cross-links unfold under physiological 
stresses, as sketched in Figure \ref{fig.one}b, and that such unfolding is 
the source of microscopic stress relaxation causing weak power-law 
rheology.  Simple estimates  \cite{envelope} suggest that physiologically
plausible stresses of order $\sim$100 Pa yield tensions sufficient to unfold 
some crosslink species. While the exponential sensitivity of 
the Bell model naturally gives rise to very broadly distributed characteristic 
unfolding and relaxation times, a broad distribution alone is not sufficient. As 
we shall see, the power-law form of the rheology requires an exponential 
distribution of cross-link tension. Just such force distributions emerge
spontaneously when simulated model networks are slowly sheared.  

\begin{figure}[t]
\begin{center}
\includegraphics[width=2.75in]{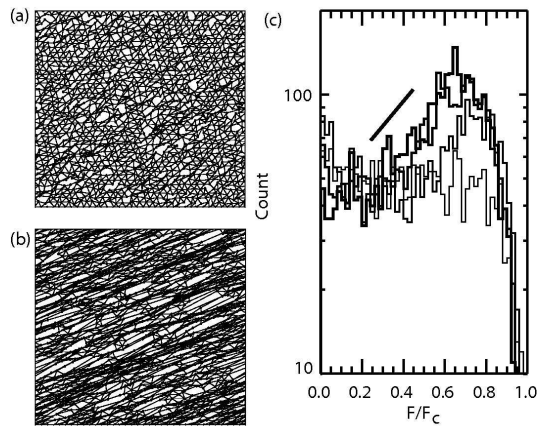} 
\end{center}
\caption{\label{fig.two} Representations of our simplified model network and
its force distribution without activated unfolding. (a) and (b) the network at 
strain $\gamma = $0.3 and 1.5 respectively. (c) Probability distributions of 
scaled force, curves for strains $\gamma = 0.3, 1.0, 1.5$, and $2.0$ 
(thin to thick) respectively. Line represents exponential behavior.}
\end{figure}

For our simulations, we constructed two-dimensional networks having 
periodic connectivity in the left/right direction, whose nodes are initially 
offset from a triangular lattice by a Gaussian-distributed amount, 
Figure \ref{fig.two}(a,b).  Force carrying mechanical links connect these nodes 
(15\% of the nearest-neighbor bonds are left empty to introduce topological 
disorder).   Networks larger than roughly 25 by 35 nodes ($\sim$$10^4$ links)
gave results that were essentially independent of system size and 
changes in network disorder.  To model serial unfolding, each link has a 
linear `sawtooth'  force-extension curve: $F = k[(x-x_0) \bmod x_c]$, where 
$k$ is a spring 
constant, $x$ is the instantaneous link extension, $x_c \equiv F_c/k$ is the 
critical extension for bond rupture and $x_0$ is the link length at the beginning 
of the simulation (typically 5-100 $x_c$). Our network does not contain 
separate rod-like and cross-link elements; each of our simulated links may be 
considered conceptually equivalent to a single inextensible rod of length 
$x_0$ connected in series with a cross-link having an infinite number of
unfoldable domains. Compressed links generate a compressive 
force (for $x<x_o, F= -k[(x_o-x) \bmod x_c] $), but this is not 
critical to our results. 

The network is sheared by translating the nodes clamped to the upper 
boundary in a series of small steps ($\Delta \gamma ~<10^{-3}$) at a 
constant strain rate, $\gamma(t)=\dot{\gamma}t$.  All nodes are first 
displaced according to an affine shear deformation, and then relaxed to 
mechanical equilibrium (zero total force on each node) by moving the 
non-boundary nodes using an `over-damped' steepest descent algorithm.
During relaxation, links move freely between branches of the force extension 
curve until an equilibrium configuration is reached.  This is equivalent to 
assuming that domains unfold and refold instantaneously when the tension 
reaches $F_c$ and zero, respectively.  

To model Bell-type 
thermally-activated unfolding a Kinetic Monte Carlo (KMC) algorithm is 
used.  The expected unfolding rate of each tensed ($F>0$) link 
is computed from its tension and the Bell model (rate $= 1/\tau_B(F))$. 
Consistent with the KMC algorithm, an exponentially distributed time step 
is generated which is inversely proportional to the total unfolding 
rate, and the link to unfold is selected in a rate weighted manner. 
To unfold the selected sub-critical link, its force extension curve is 
temporarily modified to be zero in the $x_c$-wide extension interval 
it occupies. The network is then relaxed, with the selected link typically 
moving into an adjacent, non-zero interval. 

It is illustrative to first consider the $T=0$ limit of the unfolding 
simulation ({\it i.e.} without Bell unfolding). In that case, the network 
configuration depends only on the strain, $\gamma$, and the network's 
geometrical parameters. For small strains, a roughly uniform 
distribution of link tension develops.  For strains, 
$\gamma >  0.4$, links accumulate at tensions somewhat 
smaller than $F_c$, Figure \ref{fig.two}(c). The distribution of link 
tensions approximates an increasing exponential function of force 
over a small range, $P(F) \sim exp(F/F_e)$, the qualitative form 
required to produce power-law rheology.  As the strain is increased 
further, the range of forces showing exponential behavior broadens, but 
maintains the same slope: $F_e/F_c \approx 0.4$. 

\begin{figure}[t]
\begin{center}
\includegraphics[width=2.75in]{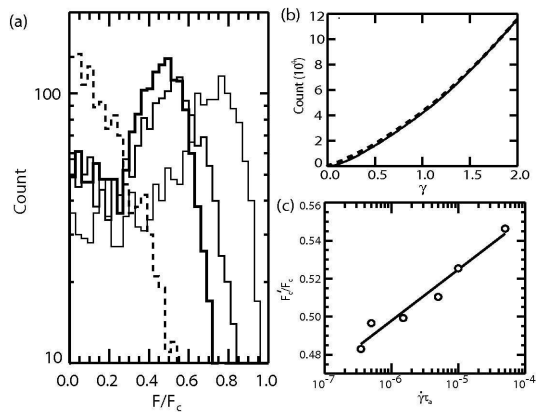} 
\end{center}
\caption{\label{fig.three} Force distributions and scaling behavior with Bell-type
activated unfolding. (a) Probability distributions of scaled force, determined for 
strain $\gamma = 1.5$, $E_B=25$ and $\tau_a=50$ nsec.  Solid lines 
are from unfolding simulations with $\dot{\gamma} = 7, 1000$ sec$^{-1}$ and the 
athermal simulation (thin to thick) respectively.  The dotted line is from 
the unbinding simulation at $\gamma= 0.5$, $E_B=25$, $\tau_a=50$ nsec, 
and $\dot{\gamma} = 10$ sec$^{-1}$  (b) The number of unfolded domains as 
a function of strain. The solid is line is during strain application and the dotted 
line is during strain reversal. (c) Most probable unfolding force as a function of 
dimensionless strain rate. 
}  
\end{figure}

In principle, the evolution of thermalized networks is more 
complicated, depending additionally on the Bell parameters $E_B$ 
and $\tau_a$ and the strain rate $\dot{\gamma}$, (assumed to be slow, 
$\dot{\gamma} \tau_a \ll 1)$. In practice, however, the qualitative 
form of $P(F)$ is quite similar to the athermal case, but with a 
somewhat smaller critical force, see Figure \ref{fig.three}(a).  Indeed, 
the $P(F)$'s from all of our thermalized unfolding simulations could be 
accurately scaled onto one another by renormalizing their forces by an
effective critical force, $F_c'$. Important for later discussion, 
the number of `unfolded' domains in the network is a monotonic 
function of network strain, with little dependence on strain rate or 
hysteresis upon strain reversal, Figure \ref{fig.three}(b).  

The effective critical force behavior has a simple explanation. Bell-type 
molecules subjected to a constant force loading rate have a 
well-defined most probable force for unfolding, which is a logarithmic 
function of the loading rate \cite{riefa}.  The constant strain rate, 
$\dot{\gamma}$, of our model slowly stretches links, leading to a  
constant force loading rate.  We can identify the resulting most probable
link unfolding force with $F_c'$, which, as expected, scales 
logarithmically with $\dot{\gamma}$, Figure \ref{fig.three}(c). Links with 
tensions even slightly smaller than $F_c'$ have an almost negligible 
Bell unfolding rate (since $E_B\gg1$), leading $P(F)$ to evolve 
much as an athermal model with `all or nothing' unfolding at critical 
force $F_c'$.

\begin{figure}[t]
\begin{center}
\includegraphics[width=2.75in]{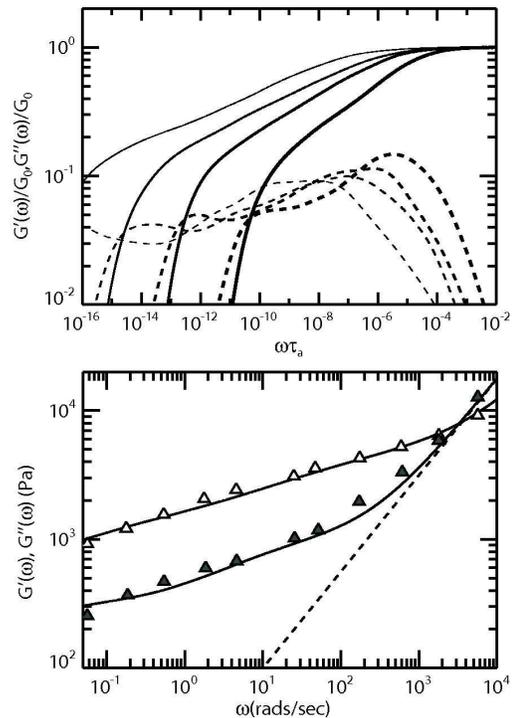} 
\end{center}
\caption{\label{fig.four} The computed frequency dependent mechanical
response of our model. (top) Dimensionless differential storage (solid)
and loss (dashed) moduli computed for $E_B = 25,30,35,40$ (thick to thin).
(bottom) The experimental data of Fabry {\it et al.} for untreated HASM cells, 
compared to our model (solid lines).  At high frequencies, a $G^* \sim 
\omega^{0.75}$, contribution was added (dashed line).  Model rheology 
generated using $E_B = 25$,  $G_o=5$ kPa, $ \tau_a= 70$ nsec and 
$\dot{\gamma}=7$ sec$^{-1}$. } 
\end{figure}

The frequency-dependent mechanical response of our network can be 
estimated by superposing Maxwell modes \cite{winter}.  Formally,  such 
a modulus describes the network's differential response to a small 
oscillatory applied stress superposed on the network's static stress.  This is 
compatible with the emerging view that cell measurements actually 
report such a differential shear modulus at a cell-generated 
prestress \cite{gardel-PNAS}.   As expected, the computed shear 
modulus, Figure \ref{fig.four}(top), has a roughly power-law form, 
$|G^*(\omega)|\sim\omega^\beta$, over a finite frequency range, with
a non-universal exponent $\beta \approx 2.5 F_c / F_c' E_B $. The 
modulus has a terminal mode at frequencies below the zero-force unfolding 
rate, $\tau_a^{-1} \exp(-E_B)$, and approaches a plateau value, $G_o$ at 
frequencies higher than $\omega_{max} \approx 1/\tau_B(F_c')$.  
Surprisingly, the frequency range with power-law rheology varies quite 
slowly with the strain rate: 
$\omega_{max} \sim \dot{\gamma}^b \tau_a^{b-1}$, with $b \approx 0.18$. 

In a separate study of forced unbinding, (rather than unfolding), 
force-extension curves with a single period were used ($F = k(x-x_o)$ for 
$|F|<F_c$, zero otherwise). Any link at zero force in mechanical equilibrium 
after a time/strain step corresponds to a link that has exceeded its critical 
force and unbound.  We then reset its $x_o$ 
parameter to its new equilibrium length to simulate rapid rebinding at zero 
force with the same network connectivity.  Activated unbinding is handled 
as before, with the KMC-selected link having its force set to zero, 
followed by network relaxation and resetting of the $x_o$'s of any 
unbound links. Unlike the unfolding simulations, the unbinding 
simulations never showed any accumulation of near critical links, 
Figure \ref{fig.three}(a), indicating such models do not produce 
power-law rheology.

Before proceeding, some discussion of the model's strain rate, 
$\dot{\gamma}$, is in order.  We suppose that in cells $\dot{\gamma}$ is 
caused by molecular motor sliding or filament treadmilling at a roughly 
constant velocity.  Rather than a uniform, pure shear deformation on the
cellular scale, a spatially random deformation field with a typical, 
mesoscopic strain rate $\dot{\gamma}$ would serve just as well. 
We further suppose that other active processes continuously 
`remodel' the network with a turnover rate comparable to $\dot{\gamma}$, 
such that typical network segments are strained to 
$\gamma \sim 1$ in a dynamic steady state that replicates the 
non-steady behavior of our model at $\gamma \sim 1$.  To be
consistent with 100-1000 second estimates for cytoskeletal turnover, 
$\dot{\gamma}$ should thus be in the range $10^{-3}$-$10^{-2}$ 
sec$^{-1}$.

The computed rheology reproduces the cell response qualitatively, but
falls short of quantitative replication.  The experimental literature 
reports at least five frequency decades of power-law scaling with exponents 
in the range $0.10 < \beta < 0.25$, while our model only yields five 
decades for $\beta < 0.17$ (or higher exponents over a narrower
frequency range).  This is due to the modest `height'
of our exponential force distribution, $P(F_c')/P(F$=$0) \approx 3$, as
in Figure \ref{fig.three}(a). Recent athermal simulations using a more realistic 
network structure yield a more pronounced exponential 
$P(F)$ \cite{didonna-preprint}, suggesting our limited frequency range may 
be an artifact of our simplified network geometry. To estimate the
model parameters corresponding to the physical case, we dimensionalized
our model, added a high-frequency contribution $G^* \sim \omega^{0.75}$, 
and compared it to literature data \cite{fabry}.  Our model reproduces the 
response of normal HASM cells, Figure \ref{fig.four}(bottom), with physically 
plausible parameter values: $E_B=25$, $\tau_a=70$ nsec and $G_o=5$ 
kPa. The simulated strain rate, $\dot{\gamma} \approx 7$ sec$^{-1}$, is higher 
than our earlier estimate; smaller dimensionless strain rates were too
computationally intensive to be accessible.  Because of the slow variation 
of model rheology with $\dot{\gamma}$, however, a more realistic value 
$\dot{\gamma} = 10^{-2}$ sec$^{-1}$, should yield a similar response 
with slightly smaller exponent, $\beta$$\approx$$0.15$, and narrow the 
frequency range with power law rheology by $(700)^{0.18} = 3.2\times$, 
or half a decade.

The arrangement of structural molecules on the cusp of conformational change 
suggests an optimal configuration for a chemical mechano-sensor---suggesting 
that cells may maintain a metabolically costly dynamic cytoskeleton as much 
for its sensory as its structural functions. Unlike other proposed sensor 
mechanisms \cite{janmeyc,tamada} that transduce molecular stress, the 
earlier noted correspondence of unfolding and network strain indicates 
the sensing of deformation on the supramolecular scale. Many cross-linking 
species specifically bind a number of signaling proteins, including heat shock 
proteins, protein kinase C, Ral A, PIP2, PIP3, PI3-kinase, and MEKK1 
(for reviews \cite{stossel,otey}).  Any of these proteins that specifically 
binds (or unbinds) cross-link domains upon forced unfolding would then
transduce the shear or extensional strain of the network, which is 
presumably a prerequisite for shape or matrix compliance sensing.  Furthermore,
the myriad, multi-domain cross-link proteins and their isoforms localized to 
different parts of the cell suggest that each different cell sub-structure could 
have its own mechano-sensing capability.  

Our cross-link unfolding model reproduces the cell response and 
makes biochemically testable predictions: that some cross-link species
should be partially unfolded under normal physiological conditions, and that
their unfolding should increase with cell deformation.  Such biochemical studies, 
combined with rheology simulations having more realistic network structure, 
hold the prospect of a cytoskeleton model grounded in polymer and 
single-molecule biophysics, one that can be integrated with mechano-sensory 
signaling pathways. 

We thank A Bausch, B DiDonna, JJ Fredberg, M Gardel,
A Levine,  T Lubensky, P Janmey and D Weitz, for useful conversations.  
Support came from the David and Lucile Packard Foundation, the Bourse 
Lavoisier and Penn's Ashton Fellowship.

$^\dagger$Present Address: Laboratoire de Spectometrie Physique,
Universit\'e Joseph Fourier, 38402 St Martin d'H\`eres, France.



\begin{thebibliography}{26}
\bibitem[{env()}]{envelope}
\bibinfo{journal}{A stress of 100 Pa = 100 pN/$\mu$m$^2$. In a gel of short 
filaments with $\sim$30 cross-links/$\mu$m$^3$, tension would be carried 
through $\sim$10 (or $30^{2/3}$) cross-links/$\mu$m$^2$.  The resulting
10 pN tension can unfold spectrin domains, and is about one quarter that 
required to unfold IgG domains.}
\end{thebibliography}


\newpage

\end{document}